\documentclass[prb,aps]{revtex4}

\usepackage{amsmath,amssymb}
\usepackage{subfigure}
\usepackage{graphicx}

\usepackage{amsfonts}
\usepackage{amsmath}
\usepackage{amssymb}

\newcommand{\be}{\begin{equation}}
\newcommand{\ee}{\end{equation}}
\newcommand{\bea}{\begin{eqnarray}}
\newcommand{\eea}{\end{eqnarray}}

\newcommand{\ket}[1]{\left| #1 \right\rangle}
\newcommand{\bra}[1]{\left\langle #1 \right|}

\newcommand{\ot}{{\,\otimes\,}}

\usepackage{graphicx}

\begin{document}

\widetext

\title{Violation of the ``information-disturbance relationship" in finite-time quantum measurements}

\author{A. Thilagam}
\email{thilaphys@gmail.com}

\affiliation{Information Technology, Engineering and Environment, \\ 
University of South Australia,  Adelaide 5095, South Australia}
\begin{abstract}
The effect of  measurement  attributes (quantum level of precision,  finite duration) 
on the classical and quantum correlations
is analysed for a pair of qubits immersed in a common reservoir.
 We show that the quantum discord is enhanced as the precision of the  measuring 
instrument is increased, and both the classical correlation and 
the quantum discord  experience noticeable changes 
during finite-time measurements  performed on a neighboring partition
of the entangled system.
The implications of these results on the ``information-disturbance relationship" are examined,
with critical analysis of the delicate roles played by  quantum non-locality  
and  non-Markovian dynamics  in the   violation of this  relationship,
which appears surprisingly  for a range of measurement attributes.
This work highlights  that the fundamental limits of  quantum mechanical measurements can be
altered by exchanges of non-classical correlations such as the quantum discord with external sources, which has
relevance for cryptographic technology.
\end{abstract}
 
\maketitle

\section{Introduction}\label{c1a}

The role of measurements in  quantum 
 correlation and decoherence processes presents 
a challenging area of investigation 
of quantum systems \cite{von,zeh,ob,Misra,FacJ,brakha,menbk}.  
Some progress has been made in the understanding of the links
between quantum measurements  and  decoherence  processes which results
in the breakdown of phases in the superposed states \cite{zuro}. In the  ``decoherence
scheme", a quantum system which survives while in 
 contact with an environment is guided to its pointer
states \cite{zurek8}. The view that  an  open quantum system is equivalent to a
system that is continuously measured by its environment has been
examined using various  approaches in other works \cite{schloss,koro}. 
Some issues still remain in relation to the  links  between  quantum correlations
 and non-locality. This  stems from difficulties in  formulating  a rigorous definition 
for non-locality, partly due the non-objectivity-non-locality issue.
In general,  quantum theory  presents a well-defined platform
in which to investigate quantum measurements,  with no apparent conflict
if the notion that  wave functions  do collapse is excluded in
favor of quantum state reduction  \cite{engl}.

Quantum measurements have been used in the
formulation of several measures of quantum correlations, including
 the quantum discord  \cite{zu,ve1,ve2}, an  entity  that is 
known to possess more generalized properties than other well
established measures such as the Wootters concurrence \cite{woot}.
The quantum discord is useful  in differentiating processes 
which are  based on locally accessible correlations from those that incorporate  generically non-classical features
 \cite{use1,maz,pii,cil,use2}.
The quantum discord of a specific system is obtained by performing
a set of positive-operator-valued measurements (POVM) 
in a neighboring partition.
At zero  quantum discord,  a  measurement
procedure allows  external observers to obtain all 
information about a bipartite system without disturbing it.

Recently, the tradeoff between  information gained due to quantum
measurement and the disturbance  on the observed quantity,
 has been examined in several works \cite{sacchi,macco,arian,buscemi,macco2}.
One study \cite{macco} showed that 
an  informative measurement  affects at least one state of the system, and
the quantity  of disturbance on the state is lower bounded 
by the amount of information  that can distinguish the input and corresponding output states.
The trade-off between the magnitude of information 
obtained via quantum measurements and disturbance on the 
evolution of the system could be  reinterpreted using 
the Heisenberg uncertainty principle  \cite{buscemi,macco2,arian}.
Heisenberg \cite{heisen} first raised the idea that any attempt to measure the position of a particle with higher precision
will result in a greater disturbance as quantified by the mean square
deviations of the momentum measurements.
In a recent experimental study \cite{rozema} involving  weak measurements,
 Heisenberg's ``measurement-disturbance relationship"
was noted to be violated. In another work \cite{mitch} involving
``entanglement-enhanced measurement",  the spinning of electrons in  atoms was observed without any disturbance to the  atomic cloud.

This  study is aimed at  examining the
attributes of the  measurement process  and  imperfections
which  distort the quantum correlations present in entangled
systems.  Following the approach in an earlier work \cite{thil},
we focus on two key attributes: a) the measurement duration,
and, b) the quantum level precision for a  model system of 
a qubit pair immersed in a common reservoir. 
In order to keep the numerical analysis tractable, 
 we adopt the Feynman's path integral framework \cite{Feyn1,Feyn2} to 
 interpret quantum measurements. In particular we employ a variant of this
formalism based  on the restricted path integral formalism \cite{menbk,mensky}. 
Within the restricted path framework,  the  continuous measurement 
of a quantity with a given  result is monitored by 
constraints imposed on the Feynman's path integral.
Accordingly an anti-Hermitian term is 
added to the Hamiltonian that describes the dynamics of the measured system.
In the presence of non-Hermitian terms introduced
 during highly precise  measurements,  the 
 measured system may evolve via one or more
 complex routes,  and the  final readout  becomes ill-defined.
The key feature in this work is thus the perusal of the idea
that state reduction can be associated with
imperfect measurements, which departs  from  conventional treatments.

This paper is organized as follows. In Section \ref{path} we describe the restricted path integral approach for energy measurements incorporating the non-ideal  attributes of the measuring device.  Description of  the quantum 
discord measures and  details of  the qubit pair system under study is provided in  Section \ref{dis}, including an analysis of the influence of critical parameters in the violation of a Bell inequality  associated with the quantum state of the  qubit pair.
 In Section \ref{meas},  the effect of the measurement precision and finite time duration on the   quantum correlation measure is evaluated and numerical results are presented. The information-measurement precision trade-off
relations is examined in Section \ref{tradeo}  and the implication of  results 
obtained in Section \ref{meas} on the ``information-disturbance relationship" is discussed. In Section \ref{mar} non-Markovianity
as quantified by the fidelity difference is used to  analyze the flow of information 
during quantum measurements,  and the issues of  quantum non-locality  
and  non-Markovian dynamics  during  violation of the ``information-disturbance relationship", is examined. The conclusion is provided in Section \ref{con}.

\section{The  restricted path integral approach for energy measurements}\label{path}
We recall the  two key elements in Feynman's path integral 
formalism \cite{Feyn1,Feyn2}: the first
involves  the superposition principle which yields the  transition 
amplitude for a given  quantum process, and 
 in the absence of measurements. Under this scheme, the probability
amplitude of the transition from the initial to the final state of the system
is obtained via  summation of the  amplitudes of  all  
possible  paths which could also  interfere with each other.  
The second feature in Feynman's formalism involves  the weight 
attached to each individual paths that is included  during the summation procedure.
This weight provides a measure of contribution of each constituent path.

The restricted path integral is derived \cite{menbk,mensky,menprl,men}
from the Feynman path integral through the incorporation of  
 a weight functional within the integrand that represents the 
summation of all tracks linking the origin and destination point.
We recall that the Feynman's propagator, 
$K_{[E]}(q^{\prime},\tau;q,0)$ in the phase-space representation
 at time $\tau$ is given  by \cite{Feyn1,Feyn2}
\be
K(q^{\prime},\tau;q,0)=
\int d[q] d[p] e^{ 
{\frac{i}  {\hbar}}\int_0^{\tau}[p \dot{q} - \widehat H_0(q,p)]dt }
\label{fey1}
\ee
where $\widehat H_0$ is the Hamiltonian of
the closed (unmeasured) quantum system and $[p]$ and $[q]$ are the
 paths in the momentum  and 
configuration spaces respectively. In Mensky's formalism,
 the output of a measured quantum system 
is expressed in terms of constrained paths associated with
 a  weight functional $w_{[E]}$ \cite{menbk}. 
This  functional may assume  a  Gaussian form, with a
damping magnitude that is proportional to the 
squared difference of the observed value along  
the paths and the actual  measurement result. 
A system subjected to measurement therefore 
evolves via a propagator
which modifies  Eq.(\ref{fey1}) according to \cite{men,ono1} 
\be
\label{feye}
K_{[E]}(q^{\prime},\tau;q,0)=
\int d[q] d[p] e^{ \frac{i}{\hbar}
\int_0^{\tau}[p \dot{q} -\widehat H_0(q,p)]dt }
w_{[E]}
\ee
This relation highlights the dependence of a selected measurement output such as
$E$   for a  measuring instrument that 
incurs  an error  $E_r$ during a measurement duration, $\tau$

The sensitivity or error during
  measurements of the energy levels of a two-level system 
is known to influence inter-level transitions \cite{ono1,ono2,audre}. 
In a recent work \cite{thil}, 
 singularities known as exceptional points \cite{Heiss} are 
shown to  appear at the branch point of eigenfunctions at a critical 
measurement precision $E_r^c$.
The significance of Mensky's formalism lies in the 
inclusion of attributes of the measuring device that may influence
 the dynamics of the quantum system under observation. 
This has obvious implications for
the evaluation of the quantum discord in quantum systems, as will be shown later
in this work.
The use of the Gaussian measure, 
$w_{[E]}$=$\exp{\left\{ -{\langle(H_0-E)^2 \rangle \over \Delta E^2} \right\} }$ 
enables the effect of the measurement to be incorporated via
the  effective Hamiltonian \cite{ono1,ono2} for a two-level system
\be
\widehat H_{eff} = \widehat H_0-i{\hbar\over{\tau  E_r^2}}(\widehat H_0-E)^2
\label{measure}
\ee
where 
 $\langle ... \rangle$ denotes the time-average for the duration $\tau$
during which  measurement  was performed. As noted earlier, $E$ 
 (see Eq.(\ref{feye})) is  the selected measurement output  after a time $\tau$ and 
$E_r$ is the error made during the measurement of the energy, $E$.
 It is evident that maximization of the product $\tau  E_r$ ensures
minimal disturbance associated with the measurement process.
This product term is  linked to  the uncertainty principle, so that
a lower limit $\tau  E_r$ would ensure maximal disturbance on the monitored
system. A large error $E_r$ and duration $\tau$ apppear as key
attributes of a weak measurement.
The  finite duration  $\tau$ yields a degree of uncertainty in energy of the observed quantum system.  
We therefore consider a weak non-Hermitian term, so that
the system under observation evolves
as $i \hbar {\partial \over \partial t}\ket{\psi(t)}= 
H_{eff} \ket{\psi(t)}$. By expanding 
the state of the  system within the unperturbed 
basis states $\ket{n}$
of the unmeasured system with Hamiltonian $\widehat H_0$ as
$|\psi(t) \rangle =\sum_n C_n(t) |n \rangle$, the coefficients
$ C_n(t)$ can be determined using the Schr\"odinger equation  based on the
Hamiltonian in Eq.(\ref{measure}).

The  Hamiltonian $\widehat H_0$ of the  unmeasured  qubit
 with energies $E_1$ ($E_2$) at state $\ket{0}$
($\ket{1}$) is of the form
\be
\label{qubit}
\widehat H_0
= - \hbar (\frac{\Delta \omega}{2}\, \sigma_{z} + V(t)\, \sigma_{x}),
\ee
where the Pauli matrices $\sigma_{x} = \ket{0} \bra{1}
+ \ket{1} \bra{0}$,  $\sigma_{z} = \ket{1} \bra{1}
- \ket{0} \bra{0}$, $\Delta \omega=2(E_1+E_2)$ and the potential $V(t)$
which induces transitions between the two levels. The  perturbation potential terms
are taken to be $V_{00}=V_{11}=0$ and
$V_{01} = V_{10}^\ast = V_0 e^{i \omega (t-t_0)}$ with $V_0$ as a real number.
 The state of the  measured system,  $\ket{\psi(t)}$
evolves as \cite{thil}
\be
 \label{evolveF}
\ket{\psi(t)} = e^{-i(E_1- i \lambda_1/4) t} C_1(t)\ket{0}
+ e^{-i(E_2-i \lambda_2/4) t} C_2(t)\ket{1}
\ee
where $\lambda_1$=$\frac{(E_1-E)^2}{2 \tau E_r^2}$ and
$\lambda_2$=$\frac{(E_2-E)^2}{2 \tau E_r^2}$ for a renormalized $E_r$.

The coefficients  $C_1(t), C_2(t)$ in Eq.(\ref{evolveF}) are  obtained using \cite{thil}
\begin{equation}
 \left[ \begin{array}{c}
  C_1(t) \\
  C_2(t) \\   \end{array} \right] =
\left[
  \begin{array}{cc}
    \cos {\kappa}t-i \alpha_1 
&-i\alpha_2\\
    -i\alpha_2 & \cos {\kappa} t+i\alpha_1 
\\  \end{array}  \right]\
\left[ \begin{array}{c}
    C_1(0) \\   C_2(0) \\
  \end{array}
\right],
\end{equation}
where $\alpha_1$=$\cos \theta \sin{\kappa} t$, 
 $\alpha_2$=$\sin \theta \sin{\kappa} t$,
  $\cos \theta$=$\frac{q}{\kappa}$, 
$\kappa$=$\sqrt{q^2+V_0^2}$, $q$=$\frac{1}{2}(\omega-\Delta E+ i \Omega/2)$, 
$\Delta E$=$(E_2-E_1)$, and  $\Omega$=$\lambda_2$-$\lambda_1$. 
The qubit  states  of the monitored  system therefore 
incorporate non-Hermitian terms which are functions of the measurement attributes
\bea
\nonumber
\ket{\chi_{\rm s}(t)}  & = & e^{- \lambda_t t/4} \left(\cos {\kappa}t-i \cos \theta \sin{\kappa t}\right)
 \ket{0} \; \\ \nonumber&& - i e^{- \lambda_t t/4} \sin \theta \sin{\kappa} t\ket{1}  \\ \nonumber
\ket{\chi_{\rm a}(t)}  & = & e^{- \lambda_t t/4}  \left(\cos {\kappa}t+i \cos \theta \sin{\kappa t} \right)\ket{1} \;
 \\ \nonumber && -i e^{- \lambda_t t/4} \sin \theta \sin{\kappa t} \ket{0},
\\
\label{eq:statesEigen}
\eea
where  $\lambda_t$=$\frac{\Delta E^2}{2 \tau E_r^2}$.
For measurement procedures which introduce very large 
errors,  $E_r \rightarrow \infty$,$\lambda_1$=$\lambda_2$=$\lambda_t$=$\cos \theta$=0, 
and the  qubit oscillates coherently between the
two levels with the Rabi frequency $2 \kappa=2 V_0$ as is well known in the unmeasured
system.

For a system in which the initial state at $t=0$ is 
 $\ket{1}$ and the final state 
at time $t$ is either $\ket{1}$ or $\ket{0}$,
the probability $P_{11}$ ($P_{10}$) of the 
system to be in the state  $\ket{1}$ ($\ket{0}$)
depends on the   relation between $V_0$ and $\lambda_t$.
At  the resonance frequencies, $\omega =\Delta E$, 
the Rabi frequency $2 \kappa_0= (4 V_0^2 -(\frac{\lambda_t}{2})^2)^{1/2}$, and 
$\cos\theta = -i\lambda_t/4 \kappa_0$. 
There exists  two  tunneling regimes with  
$V_0 > \frac{\lambda_t}{4}$ ($V_0 < \frac{\lambda_t}{4}$) applicable to the coherent (incoherent) cases.
For  the  coherent tunneling  regime we obtain \cite{thil}
\bea
\label{co1}
P_{11} &=&  e^{-\lambda_t t/2}
\left[\cos{\kappa_0} t- \frac{\lambda_t}{4 \kappa_0}\sin{\kappa_0} t \right]^2 
\\ \label{co2}
 P_{10} &=&   e^{-\lambda_t t/2} \frac{V_0^2}{\kappa_0^2}\sin^2{\kappa_0} t,
\eea
where   $\lambda_t$=$\frac{(E_2-E_1)^2}{2 \tau E_r^2}$.
The total probabilities, $P_{11}$+$P_{10} \le 1$,
the loss of normalization is dependent on the measurement
precision, $E_r$ as expected. 
For the  system undergoing  incoherent tunneling, we replace $\sin[x]$ ($\cos[x]$) by   $\sinh[x]$
($\cosh[x]$). The dynamics at the exceptional point occurs at
 $\kappa_0 = 0, \; V_0 = \frac{\lambda_t}{4}$, and both regimes merge to a point
in topological space. 
The two-level system can be seen as a non-ideal dissipative quantum  system
due to its coupling to a multitude of decay states associated with the measurement
process. 

\section{Classical correlation and quantum discord}\label{dis}

Following the formulation of quantum discord in 
Refs.\cite{zu,ve1,ve2}, we express
the quantum mutual information of a composite state $\rho$ of
 two subsystems $A$ and $B$  as
$\mathcal{I}(\rho) = S(\rho_A) + S(\rho_B) - S(\rho)$
for a density operator in $\mathcal{H}_A
\ot \mathcal{H}_B$. $\rho_{A}$ ($\rho_B$) is the  
reduced density matrix associated with $A$ ($B$) and $S(\rho_i)$ (i=A,B) denotes 
the well known  von Neumann entropy of the density operator $\rho_i$, where
$S(\rho)= - {\rm tr}(\rho \log\rho)$.
The mutual information can also be written in terms of 
quantum conditional entropy $S(\rho|\rho_A)= S(\rho) - S(\rho_A)$
as 
\be
\mathcal{I}(\rho) = S(\rho_B) - S(\rho|\rho_A)
\ee
A series of one-dimensional orthogonal projectors $\{\Pi_k\}$ induced
in $\mathcal{H}_A$ results in different outcomes of
the measurement in $\mathcal{H}_B$
via the post  measurement conditional state 
\be 
\rho_{B|k} = \frac{1}{p_k} (\Pi_k \ot \mathbb{I}_B)\rho (\Pi_k \ot
    \mathbb{I}_B)
\ee
where the probability $p_k = {\rm tr}[\rho(\Pi_k\ot \mathbb{I}_B)]$
and  $\{\Pi_k\}$ denote the one-dimensional projector indexed by the 
outcome $k$. 
From the cumulative effect of the mutually exclusive measurements 
on $A$, we obtain a conditional
entropy of the subsystem $B$ based on $\rho_{B|k}$
\be
S(\rho|\{\Pi_k\}) = \sum_k p_k S(\rho_{B|k})
\ee
which is used to obtain the
 measurement induced mutual information
$ \mathcal{I}(\rho|\{\Pi_k\}) = S(\rho_B) - S(\rho|\{\Pi_k\})$.
The  classical correlation measure based on optimal 
measurements made on $A$ is obtained as \cite{zu,ve1,ve2}
\be
\mathcal{C}_{A}(\rho) = \sup_{\{\Pi_k\}} \mathcal{I}(\rho|\{\Pi_k\})
\label{class}
\ee
The difference in $\mathcal{I}(\rho)$ and
$\mathcal{C}_A(\rho)$ yields the non symmetric  quantum discord
$\mathcal{D}_{A}(\rho) =  
\mathcal{I}(\rho) - \mathcal{C}_A(\rho)$.
The discord $\mathcal{D}_{B}(\rho)$  associated with measurements on subsystem $B$ can be 
evaluated likewise. In general $\mathcal{D}_{A}(\rho) \neq \mathcal{D}_{B}(\rho)$. 
Measurements made on a neighboring partition hold the key
to determining  the classical correlation measure between the subsystems.

\subsection{A qubit pair immersed in a common reservoir}

The joint evolution of  a pair of  two-level qubit subsystems, $A, B$ 
undergoing  decoherence  in a common reservoir
is determined by a completely positive trace preserving map expressed in
the operator-sum form \cite{salle,sudar61,kraus83}
\be
\varepsilon\left(  \rho_{AB}\right)  =\sum_{i,j}\Gamma_{i}(A)\Gamma
_{j}(B) \; \rho_{AB} \; \Gamma_{i}^{\dagger}(B)\Gamma_{j}^{\dagger}(A),
\ee
where $\Gamma_{i}(A)$ ($\Gamma_{i}(B)$) is  the Kraus operator associated with the 
decoherence  process at  $A$ ($B$). For the phase flip channel in which there 
is  loss of quantum information with conservation of energy,  the Kraus
operators are given in the basis $\{ \ket{0},\ket{1} \}$  for both
 subsystems, $k=A,B$ as  \cite{maz,salle} $\Gamma_{0}(A)= {\rm diag}(\sqrt
{1-p/2},\sqrt{1-p/2})\;\otimes\;\mathbf{1}_{B}$, 
$\Gamma_{1}(A)= {\rm diag}(\sqrt{p/2},-\sqrt{p/2})\;\otimes\;\mathbf{1}_{B}$, 
$\Gamma_{0}(B)=\mathbf{1}_{A}\;\otimes\; {\rm diag}(\sqrt{1-p/2},\sqrt{1-p/2})$
 and $\Gamma_{1}(B)=\mathbf{1}_{A}\;\otimes\; {\rm diag}(\sqrt{p/2},-\sqrt{p/2})$,
The parameter  $p=1-\exp(-\gamma t)$, where
$\gamma$ denotes the phase damping rate.

To simplify the numerical analysis, 
we consider a  joint state  of the pair of two-level qubit subsystems, $A, B$ 
 in an initial  $X$-type state  with maximally
mixed marginals (${\rho} _{A(B)}={I}_{A(B)}/2,
S({\rho}_{A}(t))=S({\rho}_{B}(t))=1$). The
density matrix appears in the form
$ {\rho}(0)=\frac{1}{4} [{I}+{\sum_{i=1..3}\;
}c_{i}{\sigma}_{A}^{i}\otimes {\sigma} _{B}^{i} ]$,
where ${I}$ is the identity operator associated with the qubit pair,
${\sigma} _{A}^{i}$, and ${\sigma} _{B}^{i}$,
and ${\sigma} _{j}^{i}$ ($j=A,B$, $i=1,2,3$) are the Pauli
operators of each qubit. $c_{i}$ ($0\leq \left\vert
c_{i}\right\vert \leq 1$) are real numbers, with 
the Werner  states sharing a common $|c_{1}|=|c_{2}|=|c_3|=c$
and $c$=1 for the Bell  basis states. 
 We assume that the usual unit trace
and positivity conditions of the density operator ${\rho}$ are satisfied.
For the class of states where $|c_{1}|=|c_{2}|=c, |c_3|=c_3$,
the evolution of the joint system is described by the matrix
\be
\label{matrix1} \rho_{_{\mathrm{A,B}}}(t)=\frac{1}{4}\left(
\begin{array}{cccc}
1+c_3 & 0 & 0 & 0 \\
0 & 1-c_3 & 2 c e^{\mu^* t}&  0 \\
0 & 2 c e^{\mu t} & 1-c_3 & 0 \\
0& 0 & 0 & 1+c_3
\end{array}
\right),
\ee
where $\mu = {\left[-2 \gamma -i (\Delta \omega_A - \Delta \omega_B)\right]t}$, and
as noted earlier $\gamma$ is  the phase damping rate.
 To simplify the analysis, we have considered the
same damping rate for the two  qubit subsystems.
$\Delta \omega_i, i=$A,B denotes the difference in energy levels 
of each qubit subsystem, we consider equivalent energy levels in 
the qubit pair.
The mutual information of state $\rho_{_{\mathrm{A,B}}}$
in Eq.~(\ref{matrix1}) is evaluated using 
$\mathcal{I}\left( {\rho}_{A}:{\rho}_{B}\right) \
=2+{\sum_{i=1}^{4}}\lambda _{i}\log\lambda _{i}$
where the eigenvalues $\lambda _{i}$ of
$\rho_{_{\mathrm{A,B}}}$ are 
$\lambda _{1,2} =\frac{1}{4}(1+c_3), 
\lambda _{3} =\frac{1}{4}(1-c_3+ 2 c e^{-2 \gamma t})$
and $\lambda _{4} =\frac{1}{4}(1-c_3- 2 c e^{-2 \gamma t})$

\subsection{Influence of $c_3$ and $c$ in the violation of a Bell inequality}
The violation of the  CHSH-Bell  inequality function $\mathcal{B}$ quantifies
quantum nonlocal correlations which  cannot be created by classical
means \cite{b1,b2}.    The  CHSH inequality Bell function $\mathcal{B}$
 is $|\mathcal{B} \leq 2$, where
$\mathcal{B} = M(\vec{a},\vec{b})-M(\vec{a},\vec{b}')+M(\vec{a}',\vec{b})+
M(\vec{a}',\vec{b}')$, where  $M(\vec{a},\vec{b})$ is the correlated 
results ($\pm 1$) arising from the measurement of two qubits  in 
directions $\vec{a}$ and $\vec{b}$. The CHSH-Bell inequality is violated when $\mathcal{B}$
exceeds 2, and the correlations is considered inaccessible by any classical means
of information transfer, while for  values less than 2, 
the local hidden-variable theory satisfies the  CHSH-Bell inequality.
 
\begin{figure}[htp]
  \begin{center}
   \subfigure{\label{abell1}\includegraphics[width=5.35cm]{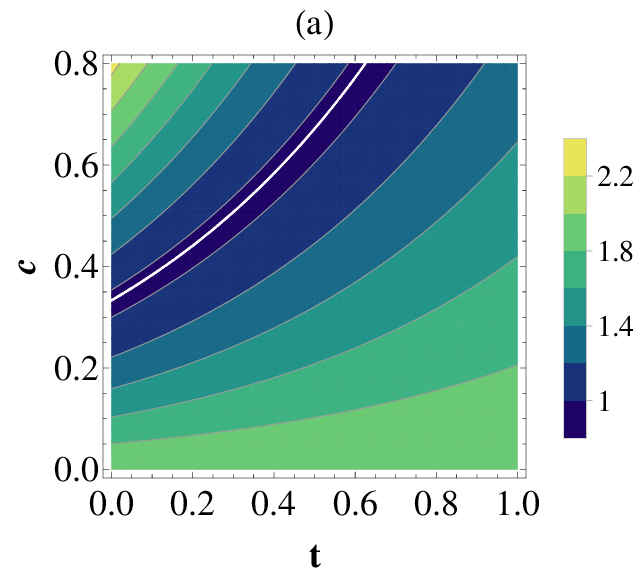}}\vspace{-1.1mm}
   \subfigure{\label{bbell1}\includegraphics[width=5.35cm]{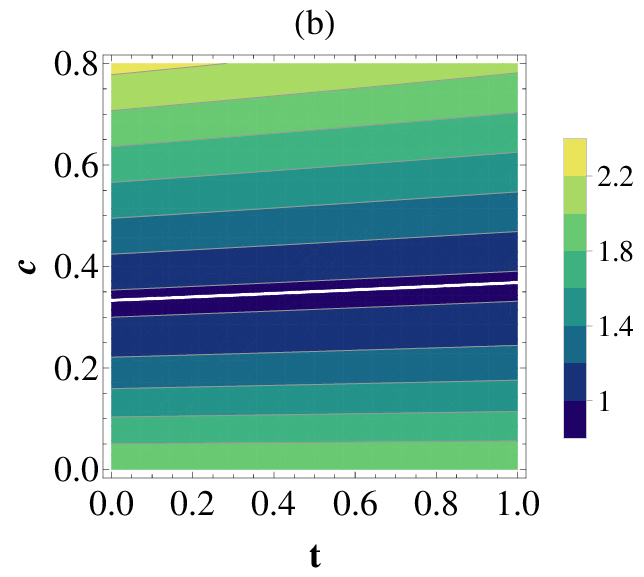}}\vspace{-1.1mm} 
\subfigure{\label{bbell1}\includegraphics[width=5.35cm]{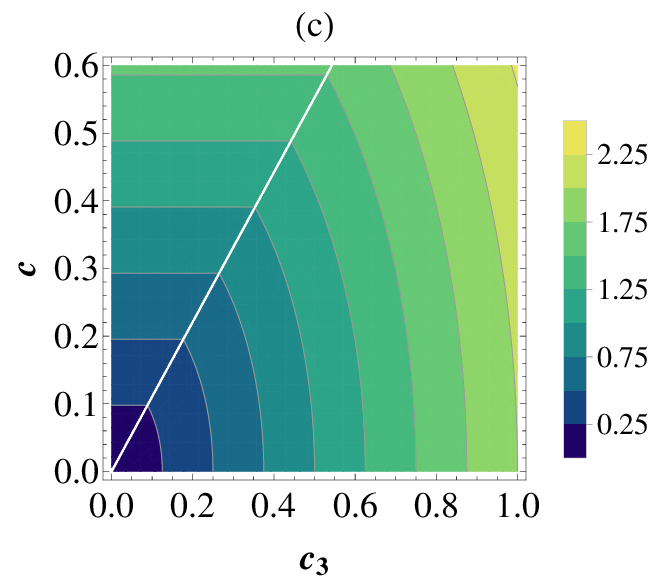}}\vspace{-1.1mm} 
     \end{center}
  \caption{(a) Bell inequality, $\mathcal{B}$ (Eq.~(\ref{b1})) as function of 
unitless $t/\tau$ ( $0 < t < \tau$)  and  $c$.  $c_3$ is fixed at $c_3=1-2 c e^{-2 \gamma t}$.
The measurement time duration, $\tau=2 \pi/V_0$=1, $c$ = 0.15, 
 and phase damping rate $g=\gamma \tau$=0.7. (b) All same except $g=\gamma \tau$=0.05.
(c) Bell inequality, $\mathcal{B}$ ( Eq.~(\ref{b1})) as function of 
  $c$ and  $c_3$, $t$=$\tau$.}
\label{bell1}
\end{figure}

We  first investigate the influence of  parameters $c_1=c_2=c$ and $c_3$  in a possible  violation of the 
CHSH-Bell inequality. These results will be compared with the   effect of $c$ and $c_3$ on classical and
quantum correlations in the next Section.
For the density matrix in  Eq.~(\ref{matrix1}),
$\mathcal{B}$  based on correlations averages, is obtained
 using the following relations \cite{bello}
\bea
\label{b1}
\mathcal{B}(t,c,c_3) &=&  {\rm Max} \; \{\mathcal{B}_1(t,c,c_3),\mathcal{B}_2(t,c,c_3) \} \\
\nonumber
\mathcal{B}_1(t,c,c_3) &=&  2 \sqrt{e^{-4 g t} c^2+{c_3}^2}\\
\nonumber
\mathcal{B}_2(t,c,c_3) &=&  2 \sqrt{2} c  e^{-2 g t}
\eea
In general, the  interplay of several parameters ($c,c_3,t,g$) makes it a  complex problem
to examine the non-locality of the two-qubit density matrix,  $\rho_{_{\mathrm{A,B}}}$.
 To simplify the approach, we note that
the eigenvalues $\lambda _{i}$ of $\rho_{_{\mathrm{A,B}}}$ in 
Eq.~(\ref{matrix1}) need to satisfy the positivity criteria of assuming only
 non-negative values. To this end, $\lambda _{4} =\frac{1}{4}(1-c_3- 2 c e^{-2 \gamma t})$
is most susceptible to violating this criteria when $c^m_3 = 1-2 c e^{-2 \gamma t}$.
Figure~\ref{bell1}a,b   show values of  $\mathcal{B}(t,c,c_3)$
as a function of  $c, t$, at two damping rates $g$, with  $c_3$=$c^m_3$.
The results indicate that  with increasing $c$, 
the system is likely to violate the Bell inequality  during the initial period of measurement.
There are subtle differences in the system non-locality arising from use of low and high $g$ as
can be inferred from Eq.~(\ref{b1}).
In Figure~\ref{bell1}c where $c_3$ is not constrained,  
 the system best exhibits  classical features at low   $c \approx 0.1$ and $c_3 \sim c$.
There is a gradual shift towards possible violation of the 
CHSH-Bell inequality as $c$ is increased, and when $\lambda _{4}$ becomes negative. 

\section{Measurement precision and quantum correlations}\label{meas}
The  classical correlation measure is obtained through all possible
local measurements on one of  the subsystems, say $A$. 
For the ideal case of a local  measurement that is instantaneous, 
 we utilize a set of orthogonal projectors 
 $\{\Pi_k=\ket{\theta _k} \bra{\theta _k}|, k=\Vert ,\perp \}$, 
which are defined in terms of the orthogonal states
\bea
\label{measureC}
\ket{\theta _{\parallel }} &=& \cos \theta 
 \ket{0} \; + e^{i\phi } \sin \theta \ket{1},\\ \nonumber
\ket{\theta _{\perp }} &=&e^{-i\phi } \sin \theta 
 \ket{0} -\cos \theta \ket{1},
\eea
where $0\leq \theta \leq \pi /2$ and $0\leq \phi \leq 2 \pi$ .
In a recent work, Galve et. al. \cite{galve} showed that orthogonal measurements
are sufficient to evaluate the quantum discord pertaining to rank 2 states
of two qubit systems, but provide tight upper bounds for higher rank (3 and 4)
states.

 Using  Eq.~(\ref{eq:statesEigen})
as a basis, we modify the projection operators in  Eq.~(\ref{measureC})
using generalized  projectors that incorporate  measurement attributes
\bea
\label{measureCm}
\ket{\theta _{\parallel }}_p &=& R(\theta)
 \ket{0} \; + e^{i\phi } S(\theta) \ket{1},\\ \nonumber
\ket{\theta _{\perp }}_p &=&e^{-i\phi } S(\theta)
 \ket{0} -R(\theta) \ket{1},
\eea
The terms $R(\theta)$= $e^{-\lambda_r t/4}
 \left(\cos {\theta}-i \frac{\lambda_r t}{4 \theta} \sin{\theta}\right)$
and $S(\theta)$=$e^{- \lambda_r t/4} \frac{\sqrt{\theta^2+ (\lambda_r t/4)^2}}{\theta} \sin{\theta}$
with  $\lambda_r$=$\frac{\Delta E^2}{2 \tau E_r^2}$, 
$\Delta E$ being the energy difference between the $\ket{1}$ and $\ket{0}$ states
of the qubit.  In the limit of $\lambda_r \rightarrow$ 0, Eq.~(\ref{measureCm}) reverts back to 
the orthogonal set in Eq.~(\ref{measureC}). 
It is implicit that the 
orthogonal measurement projections in Eq.~(\ref{measureCm})  
may be  in a state of evolution during the  measurement process.

The generalized measurements as specified by the constituent  maps  in Eq.~(\ref{measureCm})
can be projected   as follows
\bea
\label{gadg}
 &&\frac{1}{2}\;\ket{\theta _{\parallel }}_p \;  \bra{\theta _{\parallel }}_p \;+ \;  \frac{1}{2}\;
\ket{\theta _{\perp }}_p \;  \bra{\theta _{\perp }}_p,\\ \nonumber
&=&  \left(
\begin{array}{ccc}
 |R(\theta)|^2+ |S(\theta)|^2& 0\\
0 & |R(\theta)|^2+ |S(\theta)|^2
\end{array} \right ),
\\
 \nonumber
&=&   \left( \begin{array}{ccc}
1& 0\\
0 & 1
\end{array} \right ), \; \; {\tau=\lambda_r=0}
\eea
The incorporation of measurement attributes ($\tau, p$) therefore
leads to non-preservation of the trace of the density matrix in Eq.~(\ref{gadg}.
This  can be attributed to loss of the particle from the system due to the observational mapping process.
The dependence of the projected states given in Eq.~(\ref{measureCm}) on the measurement attributes,
 $\tau$ and $\lambda_r$, results in the dependence 
of the classical correlation on these same attributes which we examine  next.
 
The  reduced density matrices,
${\rho} _{B}^{(k)}$  of the neighboring subsystem $B$ 
in accordance with the two projective measurements 
($p_{\shortparallel}=p_{\bot}=1/2$) in subsystem $A$
are obtained as 
\bea
\label{matrixB1}
 \rho_{_{\mathrm B}}^{\Vert} &=& \left(
\begin{array}{cccc}
\frac{1}{2} (1-c_3 e^{- p t/2}  [\varphi_1 - \varphi_2]) & \frac{1}{2} c  e^{- 2 g t}  \varphi_1 \; \varphi_2 \\
 \frac{1}{2} c  e^{- 2 g t}  \varphi_1 \; \varphi_2 & \frac{1}{2} (1+c_3 e^{- p t/2}   [\varphi_1 - \varphi_2]) \\
\end{array}
\right),
\\
 \rho_{_{\mathrm B}}^{\perp} &=& \left(
\begin{array}{cccc}
\frac{1}{2} (1+c_3 e^{- p t/2}   [\varphi_1 - \varphi_2]) &  - \frac{1}{2} c  e^{- 2 g t}  \varphi_1 \; \varphi_2 \\
 -\frac{1}{2} c  e^{- 2 g t}  \varphi_1 \; \varphi_2 & \frac{1}{2} (1-c_3 e^{- p t/2}   [\varphi_1 - \varphi_2]) \\
\end{array}
\right).
\eea
where $\varphi_1=\cos^2\theta -\left(\frac{pt}{4 \theta}\right)^2 \sin^2\theta-\frac{\xi^2}{\theta^2}  \sin^2\theta$,
$\varphi_2=\frac{\xi^2}{\theta^2}  \sin^2\theta$  and $\xi={\sqrt{\theta^2+ (pt/4)^2}}$.
 $g=\gamma \tau$,   the dimensionless time, $t$ is obtained via  division with the measurement duration,  
$\tau=2 \pi/V_0$  and the dimensionless precision parameter, 
$p$ = $\lambda_r \tau$. The  eigenvalues of the two reduced density matrices,
${\rho} _{B}^{(k)}$  of the neighboring subsystem $B$ 
are obtained as
\bea
\label{eigenV} 
\zeta _{1,2}^{(k)} &=&\frac{1}{2}(1\pm \Theta),
\\ \nonumber
 \Theta^2&=&  c_3^2 e^{- p t}  
\left [\varphi_1 -\varphi_2 \right]^2+ 4 c^2 e^{-4 g t} e^{-p t} \varphi_1 \varphi_2
\eea
Using
Eq.~(\ref{eigenV}), we obtain
$S({\rho}_{B}^{\parallel
})$ = $S({\rho}_{B}^{\perp })$ = $- {\rm tr}(\rho \log\rho)$ = $
-\frac{1-\Theta}{2}\log_{2}
\left[\frac{1-\Theta}{2}\right]-\frac{1+\Theta}{2}\log_{2}\left[\frac{1+\Theta}{2}\right]$.
The classical correlation  given in Eq.~(\ref{class}) is  evaluated using
\be
\label{C}
\mathcal{C}(\rho)
=1- {\min }_{_{\theta,\phi}} \; R(\Theta),
\ee
Further evaluation of $\mathcal{C}(\rho)$ is simplified by the
elimination of the parameter $\phi$ due to  the  choice
of similar parameters, $|c_{1}|=|c_{2}|=c$.
The maximal value of $\Theta$ is  dependent
 on $c_3, c, \gamma$, the 
precision parameter $p$, and the measurement duration $\tau$.
It is obvious from Eq.~(\ref{eigenV}), that the influence of 
the phase damping rate becomes pronounced at  $c > c_3$.

The quantum discord is evaluated using
\be
\label{D}
\mathcal{D}(\rho) =  2+\sum_{k=1}^{4}\lambda_{k}\log_{2}\lambda_{k}
 -\mathcal{C}(\rho),
\ee
\begin{figure}[htp]
   \subfigure{\label{aa}\includegraphics[width=5.15cm]{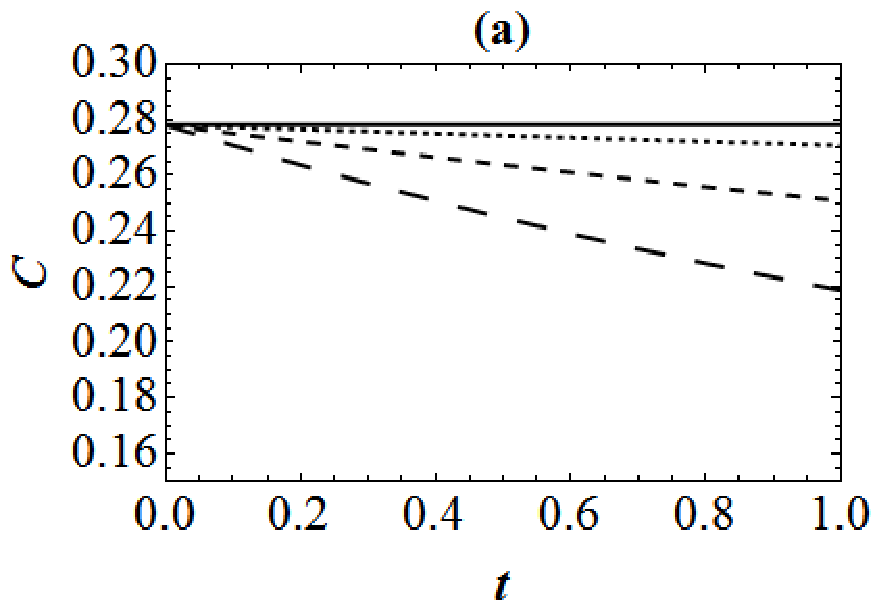}}\vspace{-1.1mm}\quad \quad \quad \quad
   \subfigure{\label{bb}\includegraphics[width=5.15cm]{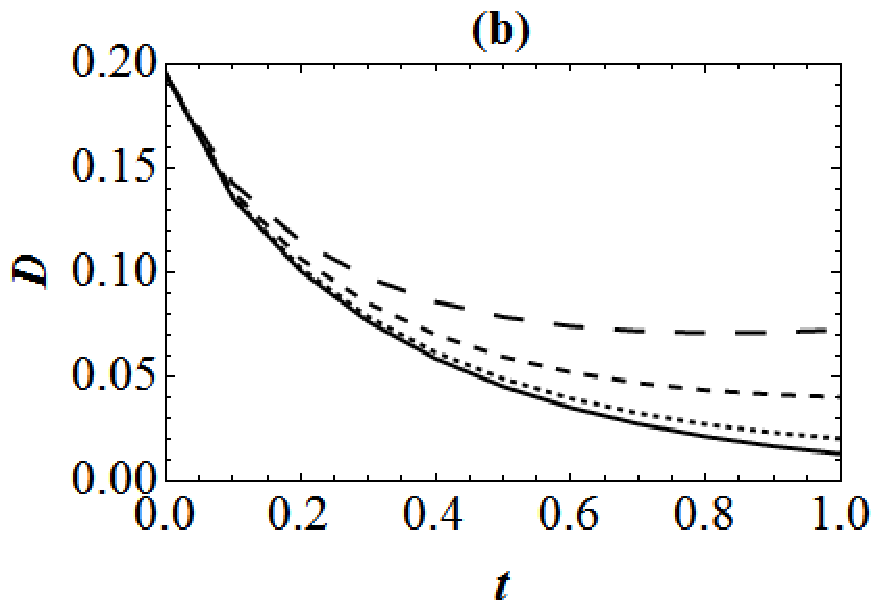}}\vspace{-1.1mm}
  \caption{(a)  Classical correlation $\mathcal{C}$ as function of $t$ ( $0<t<\tau$) 
for various values of  the measurement precision $p$. 
 $\tau$ the measurement time duration is set at 1 and $g=\gamma \tau$=0.6,
where $\gamma$ is the phase damping rate. The real numbers
$c_1=c_2=c$=0.2 and $c_3$=0.6.  The curves from top to bottom correspond
to the unitless measurement precision, $p$=0, 0.05, 0.2, 0.5.
and  (b) Quantum discord  $\mathcal{D}$ as function of $t$ ( $0<t<\tau$) 
for various values of  the measurement precision $p$=0.5, 0.2, 0.05, 0  (top to bottom).
All other parameters are the same as in (a).  
}
\label{tau1}
\end{figure}

\begin{figure}[htp]
   \subfigure{\label{aa2}\includegraphics[width=5.15cm]{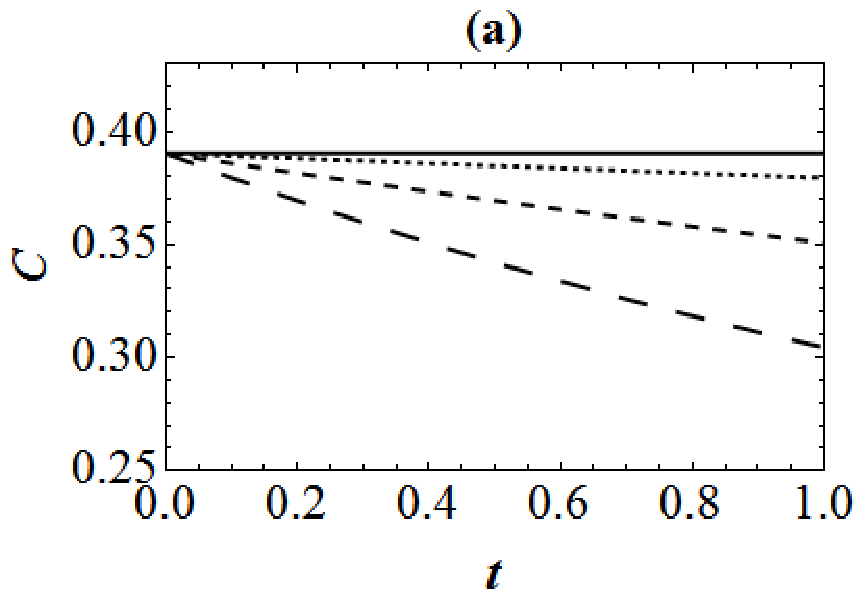}}\quad \quad \quad \quad
   \subfigure{\label{bb2}\includegraphics[width=5.15cm]{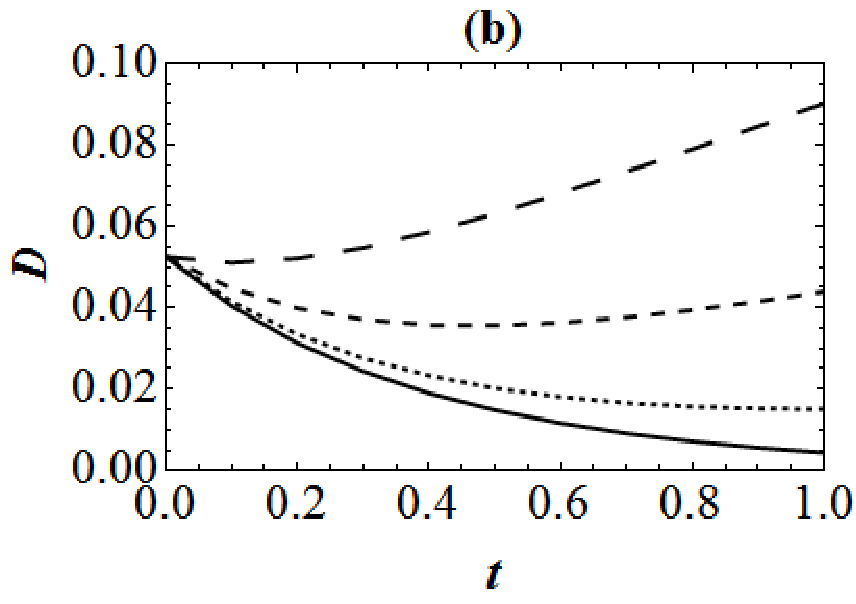}}
  \caption{a)  Classical correlation $\mathcal{C}$ as function of $t$ ( $0<t<\tau$) 
for various values of  the measurement precision $p$. 
 $\tau$ the measurement time duration is set at 1 and $g=\gamma \tau$=0.6.
 The real numbers
$c_1=c_2=c$=0.1 and $c_3$=0.7.  The curves from top to bottom correspond
to the unitless measurement precision, $p$=0, 0.05, 0.2, 0.5. and
 (b) Quantum discord  $\mathcal{D}$ as function of $t$
for various values of  the measurement precision $p$=0.5, 0.2, 0.05, 0  (top to bottom).
All other parameters are the same as in (a).  
}
\label{tau2}
\end{figure}
\begin{figure}[htp]
     \subfigure{\label{aa3}\includegraphics[width=5.15cm]{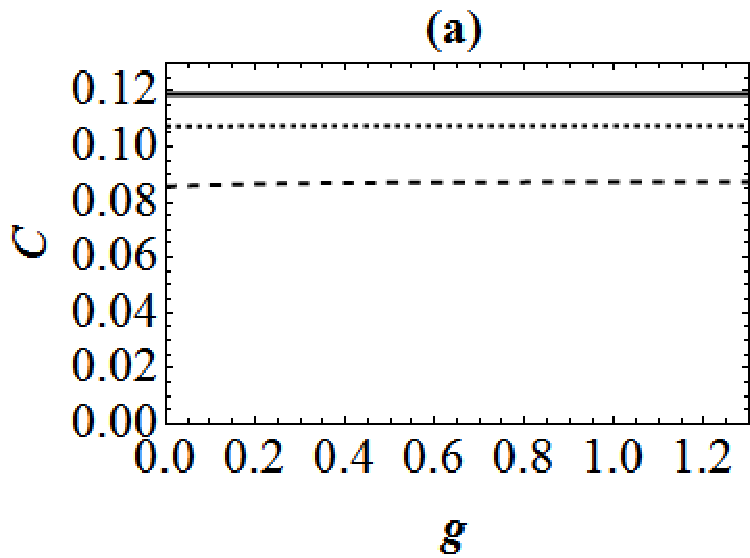}}\vspace{-1.1mm} \quad \quad \quad \quad
     \subfigure{\label{bb3}\includegraphics[width=5.15cm]{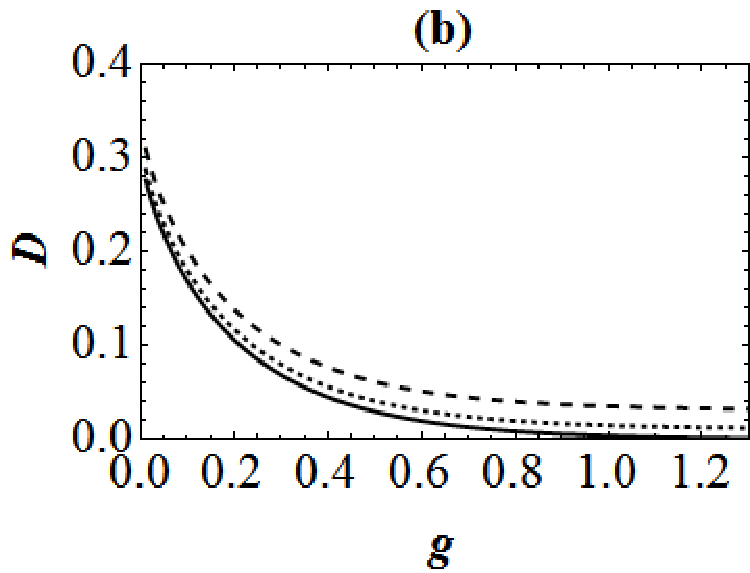}}\vspace{-1.1mm}
  \caption{(a)  Classical correlation $\mathcal{C}$ as function of  the unitless $g=\gamma \tau$
where $\gamma$ is the phase damping rate, for various values of  the measurement precision $p$ at
$t$=1. The real numbers
$c_1$=$c_2$=$c$=0.3 and $c_3$=0.4.  The curves from top to bottom correspond
to the unitless measurement precision, $p$=0, 0.2, 0.7. and (b)
  Quantum discord  $\mathcal{D}$ as function of $g$  at  $t$=1. 
for various values of  the measurement precision $p$=0.7, 0.2, 0  (top to bottom).
All other parameters are the same as in (a).  
}
\label{gauA}
\end{figure}
\begin{figure}[htp]
     \subfigure{\label{aa4}\includegraphics[width=5.15cm]{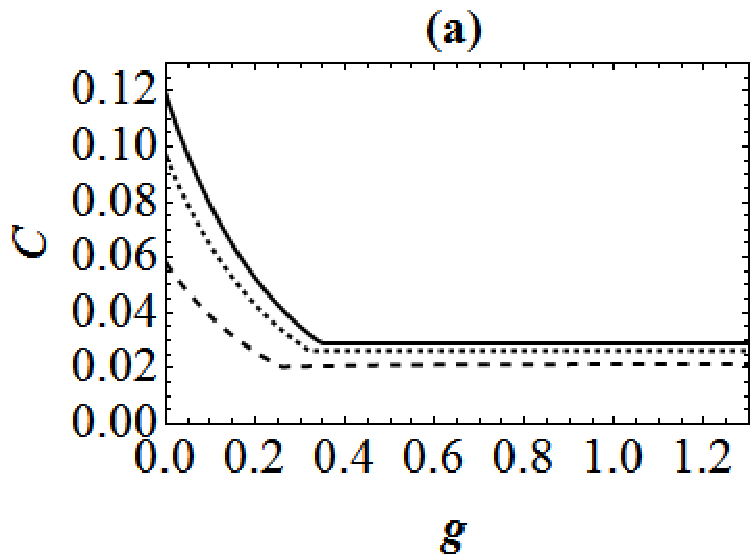}}\vspace{-1.1mm} \quad \quad \quad \quad
     \subfigure{\label{bb4}\includegraphics[width=5.15cm]{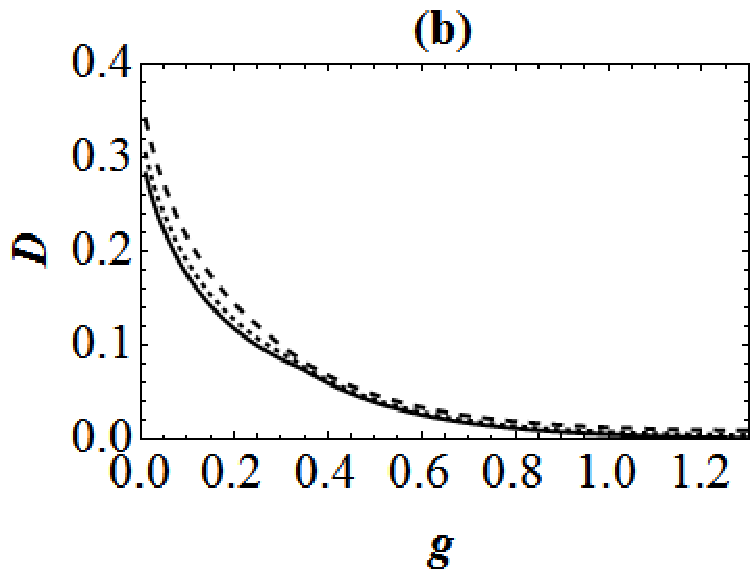}}\vspace{-1.1mm} 
      \caption{(a)  Classical correlation $\mathcal{C}$ as function of  the unitless $g=\gamma \tau$
where $\gamma$ is the phase damping rate, for various values of  the measurement precision $p$ at
$t$=1. The real numbers
$c_1$=$c_2$=$c$=0.4 and $c_3$=0.2.  The curves from top to bottom correspond
to the unitless measurement precision, $p$=0, 0.2, 0.7. and (b)
  Quantum discord  $\mathcal{D}$ as function of $g$  at  $t$=1 at
 measurement precision $p$=0.7, 0.2, 0  (top to bottom).
All other parameters are the same as in (a).  
}
\label{gauB}
\end{figure}

Figures~\ref{tau1}, \ref{tau2} highlight   changes  in the classical correlation $\mathcal{C}$ and 
quantum discord $\mathcal{D}$  as function of $t$ ( $0<t<\tau$), 
based on   numerical evaluation of Eqs.~(\ref{eigenV}), (\ref{C}), (\ref{D}) and
 the eigenvalues $\lambda _{i}$ of $\rho_{_{\mathrm{A,B}}}$ in  Eq.~(\ref{matrix1}).
$\mathcal{C}$ and $\mathcal{D}$  undergo noticeable changes due to
finite-time measurements on a neighboring partition at non-zero $p$, 
for input parameters $c_3 > c$. 
Figures~\ref{tau1}b  and \ref{tau2}b 
show that  the quantum discord $\mathcal{D}$ is enhanced, with a  corresponding decrease in 
$\mathcal{C}$ as $p$ is increased. The enhancement in the quantum discord $\mathcal{D}$ is pronounced
at a higher ratio $\frac{c_3}{c}$.  However at $c > c_3$, the numerically evaluated 
classical correlation was noted to be almost insignificant, $\mathcal{C} \approx$ 0.01 and  $\mathcal{D}$ was
seen to  be  independent of $p$. These results indicate a trend towards
more non-classical behaviour at increased $p$ for the case when $c_3 > c$, however it is not
immediately clear why a low $\mathcal{C}$  is obtained at $c_3 < c$.
Figures~\ref{gauA} and \ref{gauB} illustrate the  changes  in $\mathcal{C}$ and  $\mathcal{D}$ due to the dephasing rate $g$ for various values of  the measurement precision $p$.
At $c < c_3$, the classical correlation $\mathcal{C}$ remains independent of $g$,
a  trend  that appears only beyond a critical $g$ at $c > c_3$. This  has also been
 noted in earlier works \cite{maz,pii} for the specific case, $p$=0.
Imperfect measurements carried out 
on the subsystem $A$ therefore enhance the  non-classical correlations of
the  adjacent subsystem $B$ at $c_3 > c$.

It is not immediately clear as to the influence of quantum measurements
in the region where $c < c_3$, even though it appears that 
 there is less impact of measurements. Further investigations
are needed to confirm the role of  the  CHSH-Bell   inequality, $\mathcal{B}$ 
in relation to the $c, c_3$ parameters. For instance in 
Figure~\ref{bell1}, it appears that states which exhibit greater  classical features (at low   $c$, $c_3 > c$),
are more likely to  have reduced $\mathcal{C}$ and increased  $\mathcal{D}$ with measurements
carried out in an adjacent subsystem.
On the other hand, states which are close to CHSH-Bell   inequality violation ($c > c_3$)
seem  less influenced by imprecise measurement procedures.

\section{Information-measurement precision trade-off}\label{tradeo}

The results in Section \ref{meas} highlight  the effect of  disturbance 
on the classical and non-classical correlations of
 the adjacent reduced density matrix, ${\rho} _{B}^{(k)}$.
This  disturbance is  quantified by the
 measurement precision  $p$ and finite time duration $\tau$ associated
with the imperfect projective measurements (Eq.~(\ref{measureCm}))
on subsystem $A$. In  this Section, we examine the implications of these results on the 
``information-disturbance relationship" on subsystem $B$
as a result of imperfect measurements on  subsystem $A$.
We  note the  two sources of uncertainty ($p$, $\tau$) which give rise
to a   probabilistic distribution of the quantity being measured.

Two important measures will be used to examined the tradeoff between  information gained due to quantum
measurement and the disturbance caused during observation:  fidelity and 
trace distance. The fidelity, $F$ \cite{jozfide} 
which quantifies the distance between   two states
appear as
\be
\label{fidelity} 
F[\rho_1,\rho_2]=\left\{{\rm Tr}\left[\sqrt{\sqrt{\rho_1}\rho_2\sqrt{\rho_1}}\right]\right\}^2, 
\ee  
and  is bounded by $0\leq F[\rho_1,\rho_2]\leq 1$.
The measurement disturbance on the dynamics of a system
can be quantified  using \cite{macco} 
\be
\label{disturb} 
D_i = 1-F[\rho_1,\rho_2]
\ee
Based on the  reduced density matrices
$\rho_{_{\mathrm B}}^{\Vert}(t=0, p=0)$  and $\rho_{_{\mathrm B}}^{\Vert}(t, p)$
(Eq.~(\ref{matrixB1})), the disturbance, $D_i$ can be evaluated as a function of $t$ and $p$.

\begin{figure}[htp]
  \begin{center}
    \subfigure{\label{aa5}\includegraphics[width=5.15cm]{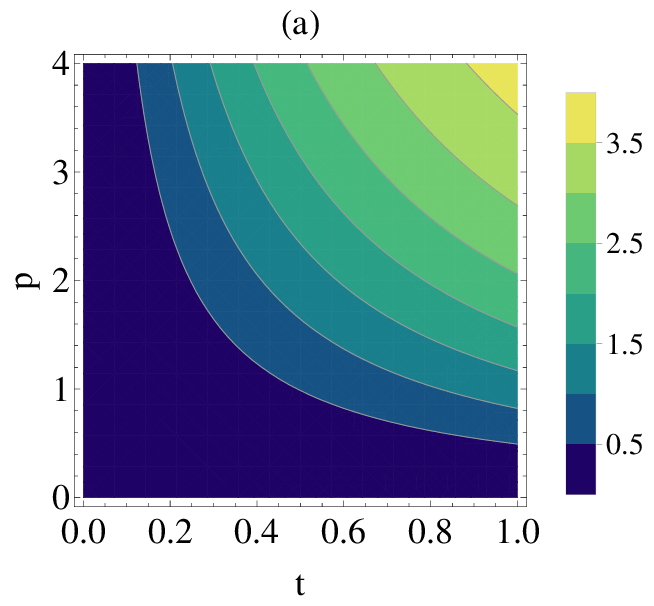}}\vspace{-1.1mm} \quad \quad \quad \quad
     \subfigure{\label{bb5}\includegraphics[width=5.15cm]{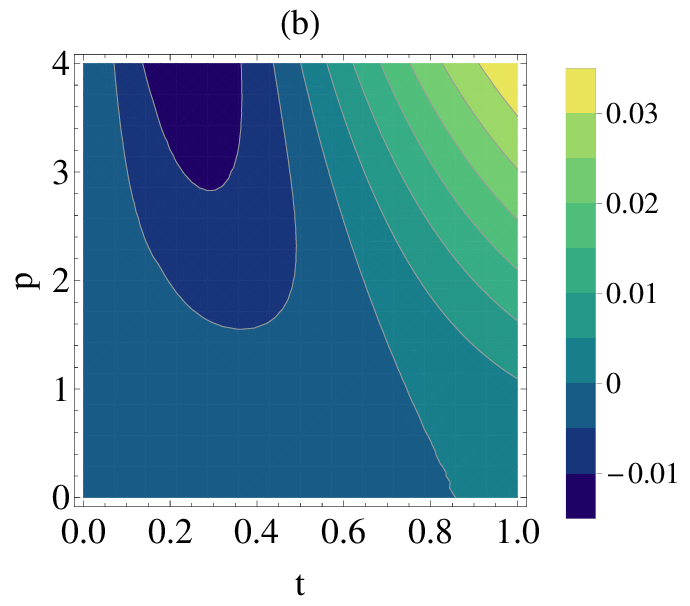}}\vspace{-1.1mm} 
     \end{center}
  \caption{(a) 
Contour-plots showing the  trade-off between the measurement disturbance
and information gain. Difference between disturbance,
 $D_i$ ($\times$ 100\%) and gain in information   $(1 - H(\nu(p,t))$ ($\times$ 100\%)
 as a function of $t$ and precision $p$.    $\rho_1$=$\rho_{_{\mathrm B}}^{\Vert}(t=0, p=0)$ 
and $\rho_2$=$\rho_{_{\mathrm B}}^{\Vert}(t, p)$ (Eq.~(\ref{matrixB1}).
$c_1=c_2=c$=0.1,  $c_3$=0.7, 
 $g=\gamma \tau$=0.5  and  $\theta$=90. 
(b) Description and parameters are the same as in (a) with the exception of  $c_1=c_2=c$=0.4,  $c_3$=0.1.
A range of $p$ and $t$ for which the difference between disturbance,
 and gain in information is negative becomes noticeable.
}
\label{macca}
\end{figure}

 To quantify the information gained from the system, we define  the  uncertainty $H(\nu)$
based on  the parameter $\nu$ where 
\be
\label{nute}
\nu = \frac{1}{2}-\frac{1}{2} T_d(\rho_1,\rho_2)
\ee
The trace distance, $T_d$ between  density matrices, $\rho_1,\rho_2$, 
 is given by half of the trace norm of the difference of the matrices as 
$T_d(\rho_1,\rho_2)$ = $\frac{1}{2} {\rm Tr}[|\rho_1-\rho_2|]$ and 
$H(x) = - x \log_2 x-(1-x) \log_2 (1-x)$. In order to 
analyse the influence of the precision $p$ and $t$, we 
consider $\rho_1$=$\rho_{_{\mathrm B}}^{\Vert}(t=0, p=0)$ 
and $\rho_2$=$\rho_{_{\mathrm B}}^{\Vert}(t, p)$.
We next investigate the
information-disturbance tradeoff relation \cite{macco}
\be
\label{trade}
1-F[\rho_1,\rho_2] \ge 1 - H[\nu(p,t)]
\ee
where the mutual information (1-$H[\nu(p,t)]$) is evaluated
based on $\nu$ (Eq.~(\ref{nute})) for given values of the  measurement attributes, $p,t$.
Eq.~(\ref{trade}) specifies that the disturbance
between two states $\rho_1,\rho_2$ has a lower bound, quantified
by the gain in information due to  measurements. 
An alternative interpretation of Eq.~(\ref{trade}) was also provided \cite{macco}
through the existence of a lower limit to the sum of the 
disturbance $1-F$ and the uncertainty $H[\nu(p,t)]$ that is
based on  verification of  the difference in the two states, $\rho_1,\rho_2$.
The difference between disturbance,  
$D_i$ ($\times$ 100\%) and gain in information   $(1 - H(\nu(p,t))$ ($\times$ 100\%)
 as a function of $t$ and precision $p$ is shown
in  Figure~\ref{macca}a,b. We note that  for the input parameters 
($c_1=c_2=c$=0.4,  $c_3$=0.1), there exists a   range of $p$ and $t$ for which the
``information-disturbance relationship" (Eq.~(\ref{trade}))
formulated in Ref.\cite{macco} is  violated.

Comparing the  results  in Figure~\ref{macca} a,b with those in 
  Figures~\ref{tau1}, \ref{tau2}, one notes  that the appearance of 
increased quantum discord is invariably linked to the non-violation of the inequality
in Eq.~(\ref{trade})  when $c_3 > c$. The difference between the disturbance, $D_i$ (Eq.~(\ref{disturb}))
and the mutual information (1-$H(\nu(p,t))$) yields a measure of the quantum discord. 
This difference is accentuated at increasing $p$, a trend that is also observed
in the enhancement of the quantum discord $\mathcal{D}$ at higher $p$.
One can expect a zero discord when the  lower bound in Eq.~(\ref{trade})
is reached, at which point the disturbance on the system equates the
amount of information that can be retrieved. 
 Another important observation relates
to the case when  $c > c_3$, where we earlier noted the existence of 
a small $\mathcal{C} \approx$ 0.01 and  $\mathcal{D}$  that was immune to changes
in $p$. Interestingly, 
we note that a violation of Eq.~(\ref{trade}) occurs when $c > c_3$ and 
for a range of $p, t$ as illustrated in Figure~\ref{macca}b.

One possible explanation of the noted
violation may lie in the  presence of other unseen neighboring  subsystems 
which gives rise to a net deficit in quantum discord, as far as the two
known subsystems $A, B$ are concerned. This results in 
greater retrieval of information than the actual disturbance on the system,
with bearings very similar to the  Maxwell's demon model \cite{Szilard}.
In the latter model,   positive entropy production during  measurement arise from work performed by
other agents, ensuring  that 
the second law of thermodynamics remains intact.

The results in Figure~\ref{macca} a,b can partly be interpreted  on the basis of 
earlier obtained results  in Figure~\ref{bell1} a,b,c, where we noted
that at higher $c > c_3$, there is trend towards violation of the CHSH-Bell inequality.
The violation of Heisenberg's ``measurement-disturbance relationship"
as evidenced by the nature of the  $c, c_3$ input parameters, 
may have its origins in non-local quantum states which 
are also influenced by  these same parameters.   
 To this end, investigations
involving rigorous mathematical formulations \cite{anand}  of the underlying
abstract Hilbert space are needed  to provide greater insight to the links
between the information-disturbance tradeoff relation, quantum discord and quantum non-locality
based on  the CHSH-Bell  inequality function $\mathcal{B}$ (Eq.~(\ref{b1})).
The  results obtained here may be useful in the interpretation of 
experimental results \cite{rozema}  showing similar violation of the ``measurement-disturbance relationship".

\section {Non-Markovianity during quantum measurements}\label{mar}
\begin{figure}[htp]
 \begin{center}
     \subfigure{\label{aar1}\includegraphics[width=3.6cm]{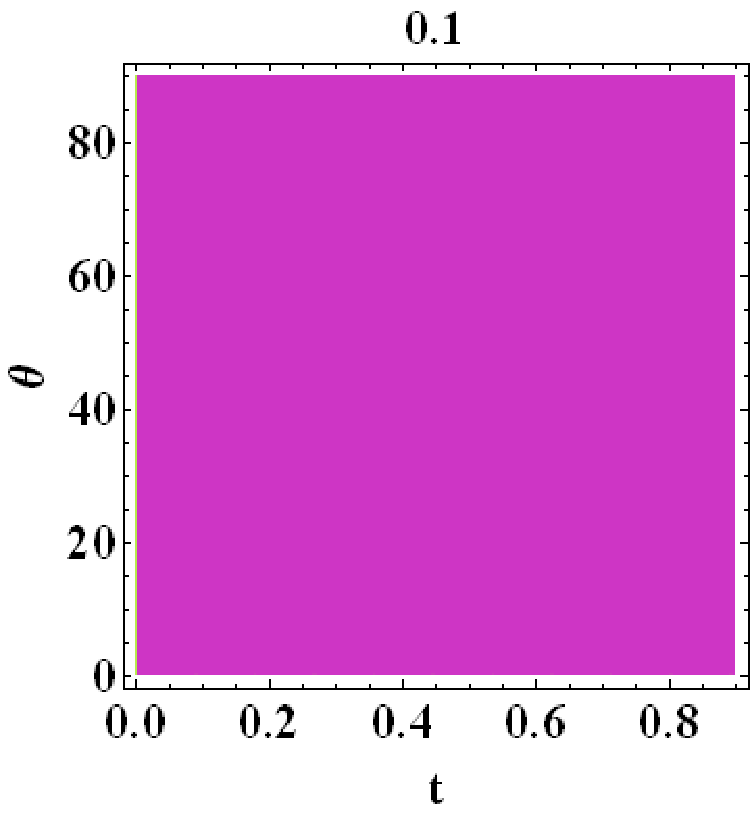}}\vspace{-1.1mm} \hspace{1.1mm}
     \subfigure{\label{bbr1}\includegraphics[width=3.6cm]{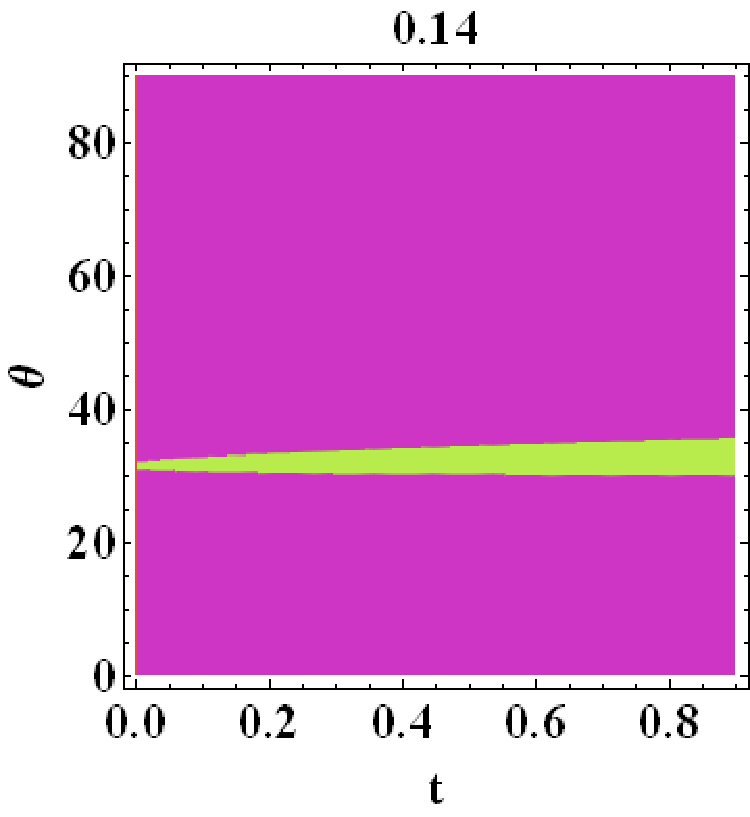}}\vspace{-1.1mm} \hspace{1.1mm}
\subfigure{\label{aar2}\includegraphics[width=3.6cm]{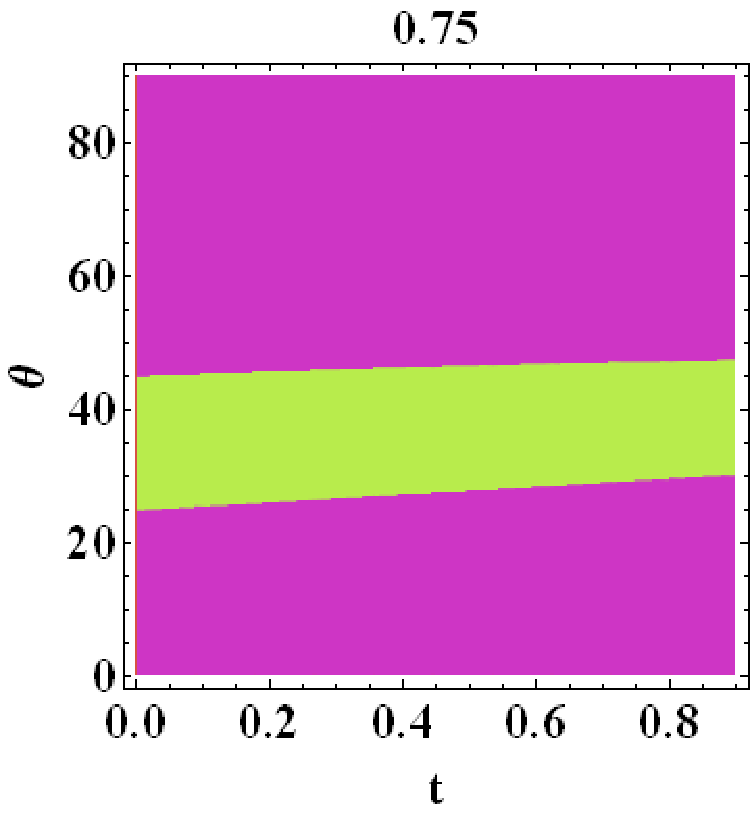}}\vspace{-1.1mm} \hspace{1.1mm}
     \subfigure{\label{bbr2}\includegraphics[width=3.6cm]{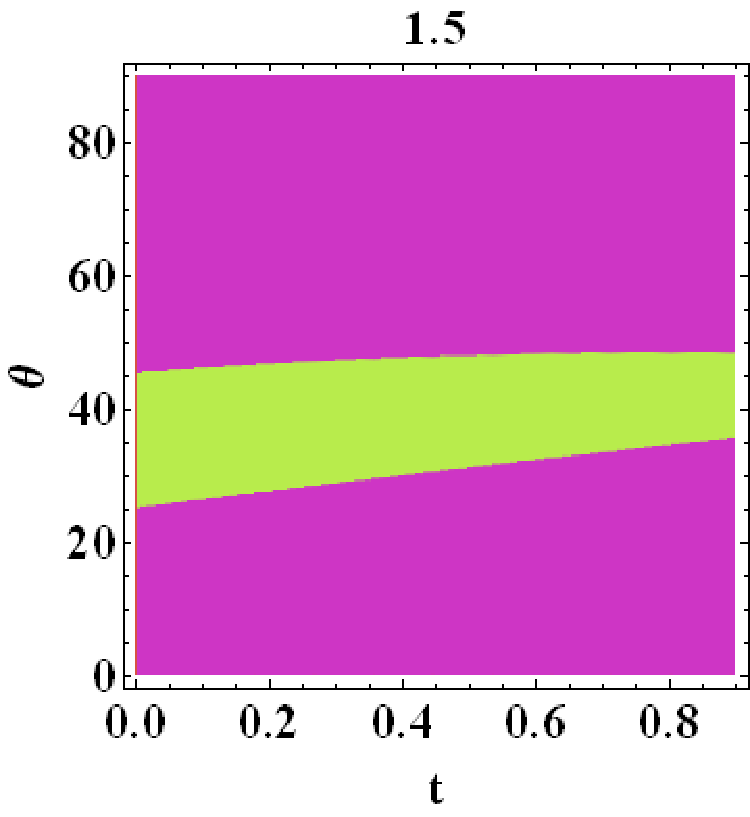}}\vspace{-1.1mm} \hspace{1.1mm}
\subfigure{\label{aar3}\includegraphics[width=3.6cm]{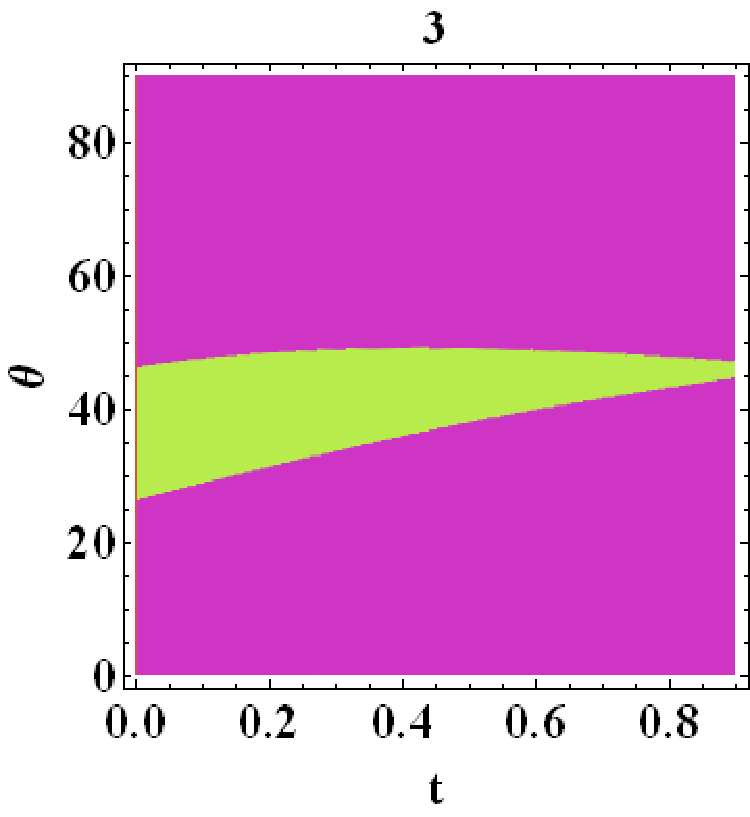}}\vspace{-1.1mm} \hspace{1.1mm}
     \subfigure{\label{bbr4}\includegraphics[width=3.6cm]{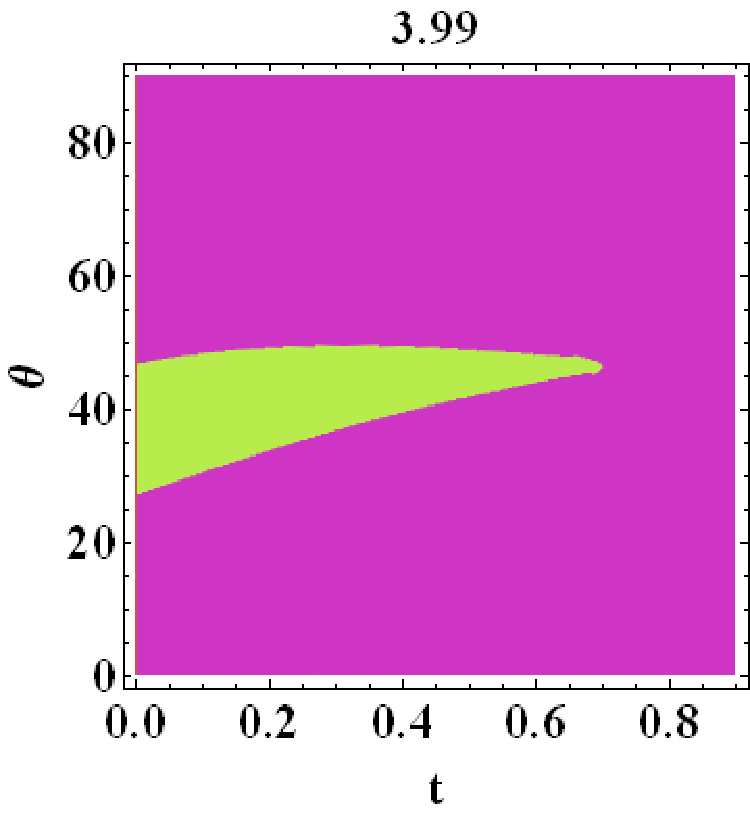}}\vspace{-1.1mm} \hspace{1.1mm}
      \end{center}
      \caption{Fidelity difference $\Delta(t, \tau)$  as a function of $t$ and $\theta$
for the density matrices corresponding  to 
 $\rho_1$=$\rho_{_{\mathrm B}}^{\Vert}(t=0, p=0)$ 
and $\rho_2$=$\rho_{_{\mathrm B}}^{\Vert}(t, p)$ (Eq.~(\ref{matrixB1}). We set
$\tau$=0.1,  $c_1=c_2=c$=0.1,  $c_3$=0.8, 
 $g=\gamma \tau$=0.1. Values of $p$ are indicated at the top of each figure.
 Regions of negative values   indicating non-Markovianity are shaded green show increase
with precision $p$, and are dominant in the range  $25 < \theta  < 42 \deg$.
 }
 \label{MakA}
\end{figure}

\begin{figure}[htp]
 \begin{center}
   \subfigure{\label{aat1}\includegraphics[width=3.8cm]{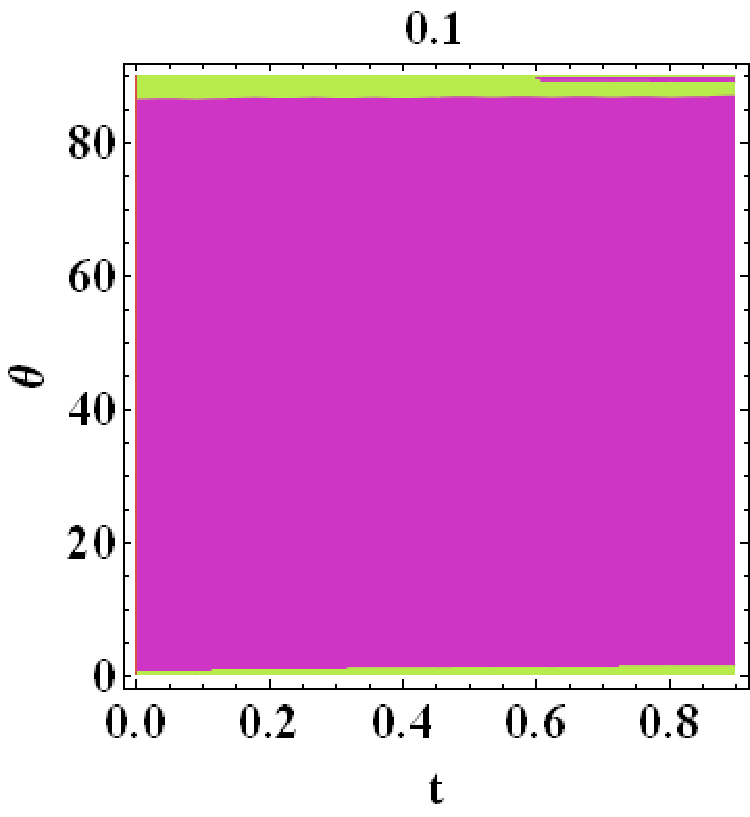}}\vspace{-1.1mm} \hspace{1.1mm}
     \subfigure{\label{bbt1}\includegraphics[width=3.8cm]{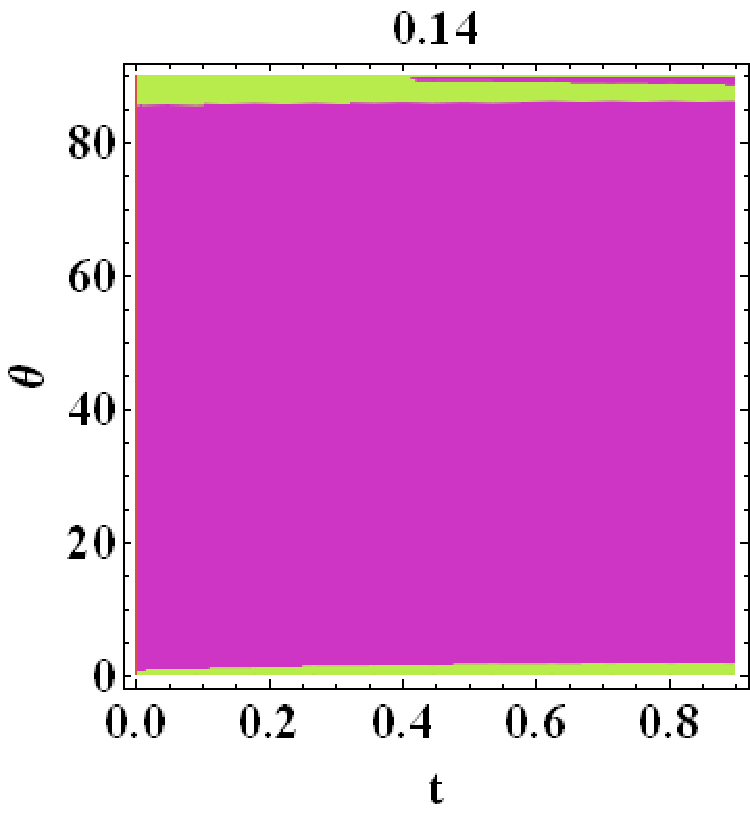}}\vspace{-1.1mm} \hspace{1.1mm}
\subfigure{\label{aat2}\includegraphics[width=3.8cm]{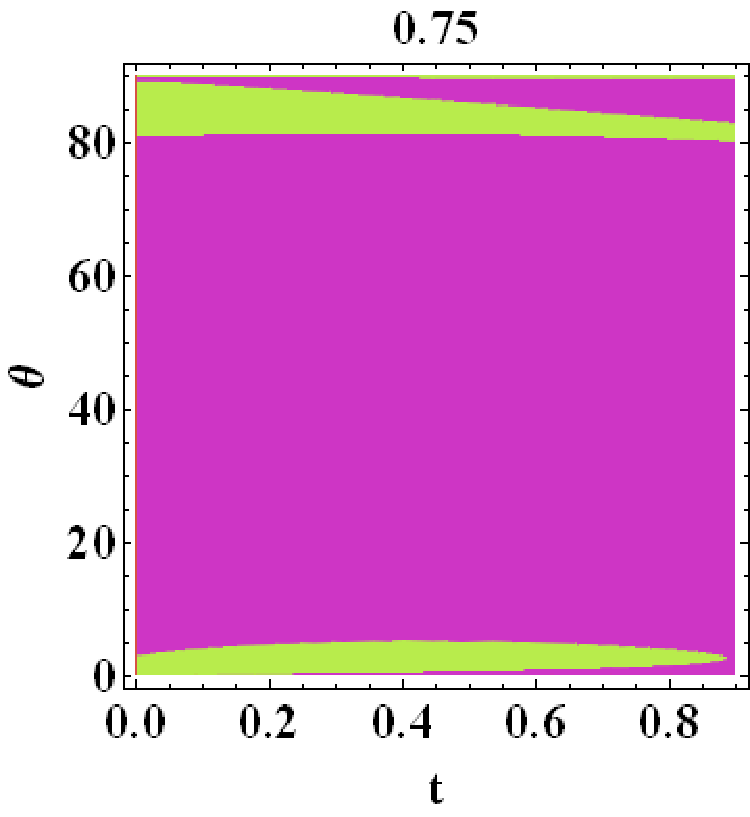}}\vspace{-1.1mm} \hspace{1.1mm}
     \subfigure{\label{bbt2}\includegraphics[width=3.8cm]{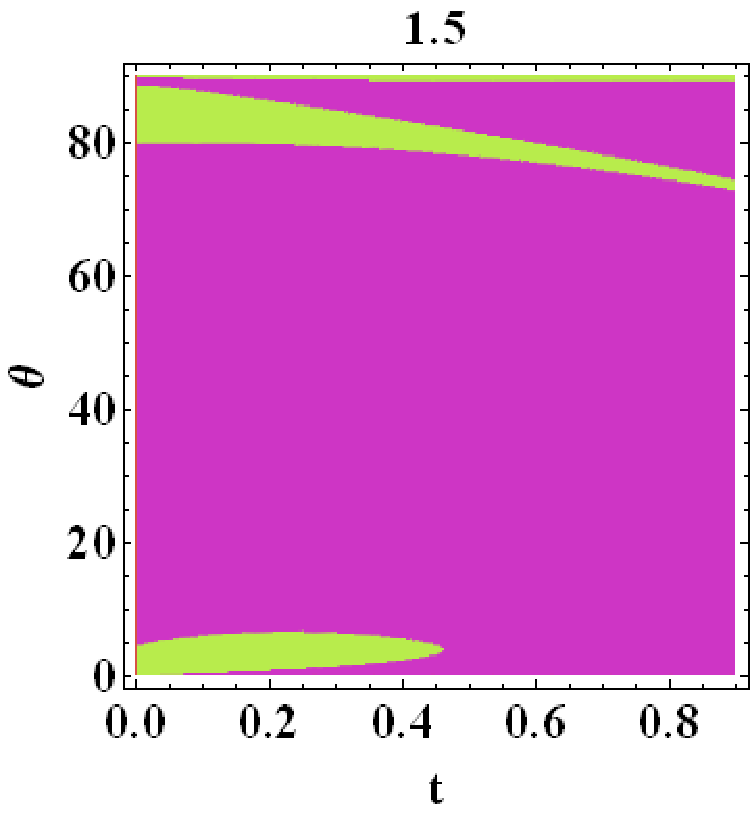}}\vspace{-1.1mm} \hspace{1.1mm}
\subfigure{\label{aat3}\includegraphics[width=3.8cm]{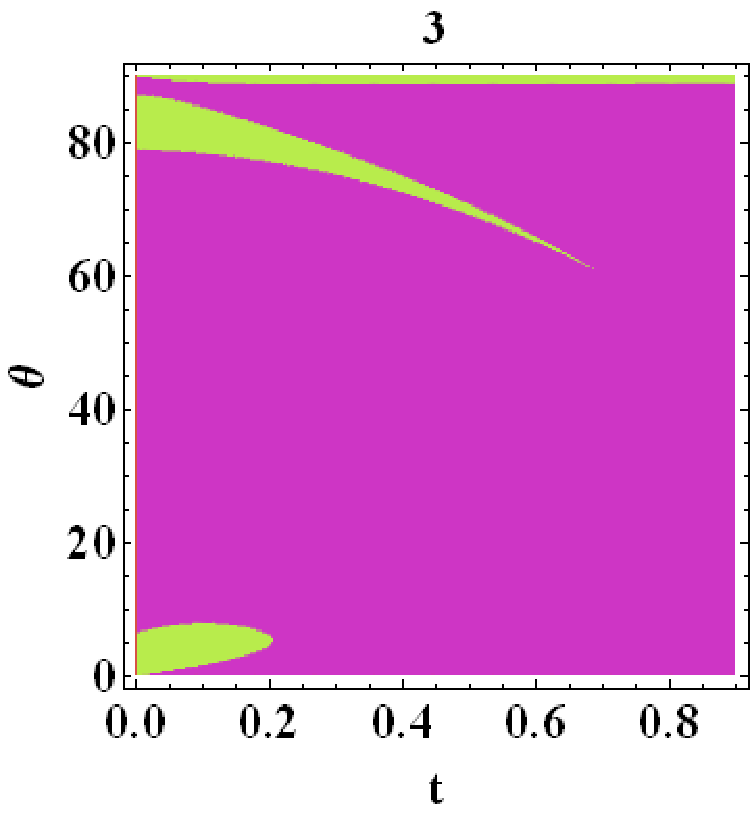}}\vspace{-1.1mm} \hspace{1.1mm}
     \subfigure{\label{bbt3}\includegraphics[width=3.8cm]{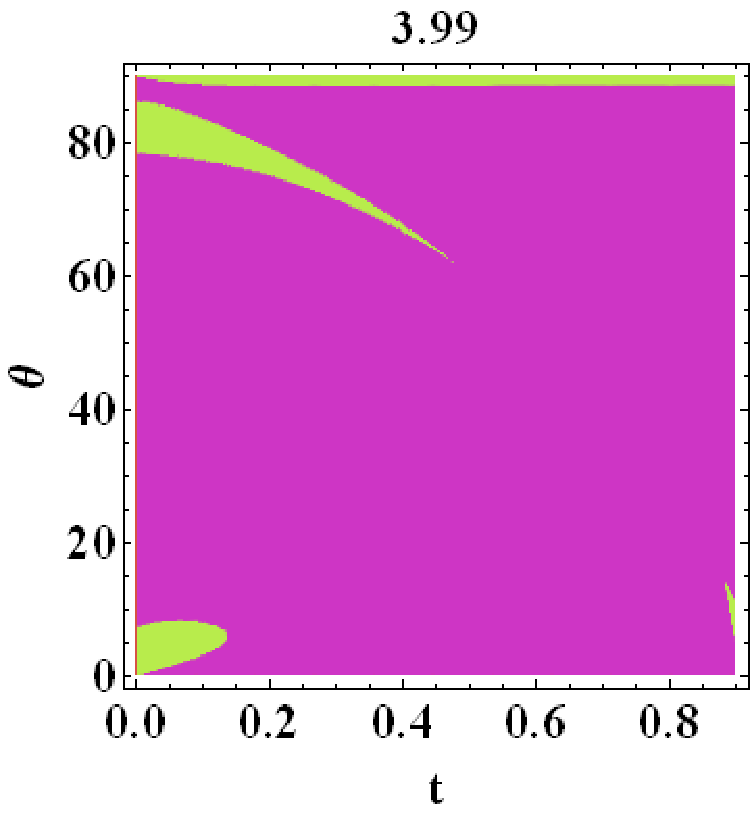}}\vspace{-1.1mm} \hspace{1.1mm}
      \end{center}
  \caption{Fidelity difference $\Delta(t, \tau)$  as a function of $t$ and $\theta$
for the density matrices corresponding  to 
 $\rho_1$=$\rho_{_{\mathrm B}}^{\Vert}(t=0, p=0)$ 
and $\rho_2$=$\rho_{_{\mathrm B}}^{\Vert}(t, p)$ (Eq.~(\ref{matrixB1}). We set
$\tau$=0.1,  $c_1=c_2=c$=0.4,  $c_3$=0.1, 
 $g=\gamma \tau$=0.1. Values of $p$ are indicated at the top of each figure.
Regions of  negative values   indicating non-Markovianity are shaded green and 
are located at the peripheral regions, $\theta  \approx 0, 90$.
}
\label{MakB}
\end{figure}

To better understand the flow of information during quantum measurements, we consider the
appearance of non-Markovianity, in  relation to the attributes, $p,t$. 
Quantum systems undergoing Markovian dynamics observe
 a completely positive, trace preserving dynamical map $\Lambda(t)$, 
$\rho(0) \rightarrow \rho(t)=\Lambda(t)\rho(0)$, which constitutes
the  one parameter semi-group obeying the composition law~ \cite{wolf,raja,breu},
$\Lambda(t_1)\Lambda(t_2)=\Lambda(t_1+t_2), \ t_1,t_2\geq 0$.
Accordingly, the fidelity function  $F[\rho(t),\rho(t+\tau)]$ involving the 
initial state $\rho(t)$ and the evolved state  $\rho(t+\tau)$ at a later time $t+\tau$,
 under Markovian dynamics satisfies the inequality  \cite{raja}
\begin{eqnarray}
\label{ine}
F[\rho(t),\rho(t+\tau)]\equiv F[\Lambda(t)\rho(0),\Lambda(t)\rho(\tau)] \nonumber \\ 
\Rightarrow F[\rho(t),\rho(t+\tau)]\geq F[\rho(0),\rho(\tau)].
\end{eqnarray}
Any violation of this inequality is a  signature of non-Markovian dynamics  which can be observed
via  the fidelity difference function 
\begin{equation}
\label{fdiff} 
\Delta(t,\tau)= \frac{F[\rho(t),\rho(t+\tau)] - F[\rho(0),\rho(\tau)]}{F[\rho(0),\rho(\tau)]},
\end{equation} 
Negative values of $\Delta(t, \tau)$ serve as sufficient  but not necessary condition  of 
non-Markovianity. Using Eq.~(\ref{fdiff}),
we have evaluated the fidelity difference $G(t,\tau)$  as a function of $t$ and $\theta$
for the density matrices corresponding  to 
 $\rho_1$=$\rho_{_{\mathrm B}}^{\Vert}(t=0, p=0)$ 
and $\rho_2$=$\rho_{_{\mathrm B}}^{\Vert}(t, p)$ (Eq.~(\ref{matrixB1}),
 as illustrated in Figures~\ref{MakA} and \ref{MakB}. The figures highlight
important differences  between  systems where $c_3 > c$ and those with $c > c_3$.
In the regions midway: 
$25 < \theta  < 42 $, there is enhancement of 
non-Markovianity with precision $p$ when $c_3 > c$.
In  systems where $c_3 <  c$, the non-Markovian regions are located at the peripheral regions, $\theta  \approx 0, 90$.
We note that at $c_3 > c$ the optimized angle $\theta$ used in the
evaluation of the  classical correlation   in Eq.~(\ref{C}) is about
$90$, decreasing gradually with increase $p$.  At $c_3 < c$,
the optimized angle $\theta < 30$. These results highlight the role of $c, c_3$ parameters in determining the 
links between  non-Markovian dynamics and optimization processes associated
with the  classical correlation measure.

The findings in Figures~\ref{MakA} and \ref{MakB} may provide a speculative basis 
to  examine links between violation  of the ``measurement-disturbance relationship"  in
Eq.~(\ref{trade}) and non-Markovianity.  Is it  possible that a pathway for violation of this relationship is
 attained
when gain in information exceeds  disturbance
between two states via non-Markovian processes? This  challenging question  needs experimental
verification using more generalized systems, as numerical results related to just
two subsystems have been provided here.

\section{Conclusion}\label{con}
In conclusion, we have presented 
results of the influence of non-ideal attributes such as the
  measurement precision and finite measurement time duration
 on the classical correlation and  quantum discord
for a qubit pair immersed in a common environment.
The results show that  the quantum discord is enhanced as the precision of the  measuring 
instrument is increased for a range of parameters, and both the classical correlation and 
the quantum discord  undergo noticeable changes 
during the duration when  measurements are  performed on a neighboring partition.
We also conclude that increased quantum discord within two subsystems, 
is invariably linked to the non-violation of the inequality
associated with the ``information-disturbance relationship". We note that
a zero discord corresponds to the  lower bound in this inequality,
 consistent with the point at which  the disturbance on the system equates the
amount of information that can be retrieved. A violation of  this inequality  indicates
a deficit in quantum discord,  with the possibility
that other undetected  agents may be responsible in giving rise to 
 a greater retrieval of information than the actual disturbance on the system.
Overall, the results obtained in this work indicate that 
the fundamental limits of  quantum mechanical measurements may be
altered by exchanges involving non-classical correlations such as the quantum discord with external sources. 

The  violation of the ``information-disturbance relationship" 
may have links with quantum non-locality and 
non-Markovian quantum  dynamics of states that do not necessarily evolve via
 a completely positive, trace preserving dynamical maps.  
This study identifies (though not conclusively) that 
a possible pathway for violation of this relationship may occur
when gain in information exceeds  disturbances
between two states via non-Markovian processes.
Further scrutiny of the intricate links between these entities 
require mathematically rigorous approaches \cite{anand} and experimental 
verifications, however the results obtained in this study have wider implications for 
exploiting the subtleties of the Uncertainty Principle in multipartite systems.
Cryptographic technologies specify that eavesdroppers can be detected as a 
result of disturbances caused by their measuring
activities. The results obtained here indicates 
that eavesdroppers  can remain undetected in some instances (when positivity of the
density matrix of the observed system is violated).
These ideas may be further extended to the development of sensitive quantum probes
of fragile material systems, involving single atoms and molecules, and including living tissue
matter.

Lastly, this study  shows that the joint examination of 
several entities (non-locality, non-Markovianity, negative quantum discord) 
is needed in investigations involving
 quantum measurements of optics and nanostructure systems \cite{hall,titt}. 
The possibility that an analogous
 ``information-disturbance relationship" may be violated in 
 light-harvesting systems \cite{l1,l2,l3,13b,l4} is an area for future investigation
of systems which display  exceptionally high  efficiencies of  energy transfer processes.

\section{Acknowledgements}
This research was undertaken on the NCI National Facility in Canberra, Australia, 
which is supported by the Australian Commonwealth Government.
The author gratefully acknowledges the  support of  the Julian Schwinger Foundation Grant,
JSF-12-06-0000.  The author would like to thank the anonymous referees for
helpful comments.

\end{document}